\documentclass{ws-ijmpd}
\usepackage{epsfig}

\newcommand{\be}{\begin{equation}}
\newcommand{\ee}{\end{equation}}
\newcommand{\bea}{\begin{eqnarray}}
\newcommand{\eea}{\end{eqnarray}}
\def\simgt{\rlap{\lower 3.5 pt \hbox{$\mathchar \sim$}} \raise 1pt
Ê \hbox {$>$}}

\begin{document}
\markboth{E.I. Guendelman}
{Wormholes and Child Universes}

%
\catchline{}{}{}{}{}
%

\title{Wormholes and Child Universes}

\author{E.I. Guendelman}

\address{Physics Department, Ben-Gurion University of the Negev,
Beer-Sheva 84105, Israel}

\maketitle

\begin{history}
\received{Day Month Year}
\revised{Day Month Year}
\comby{Managing Editor}
\end{history}

\date{\today}

\begin{abstract} Evidence to the case that classical gravitation provides
 the clue to make sense out of quantum gravity is presented. The key observation is the
existence in classical gravitation of child universe solutions or "almost"
solutions , "almost" because of some singularity problems. The difficulties of these
child universe solutions due to their  generic singularity problems will be very
likely be cured by quantum effects, just like for example "almost" instanton solutions
are made relevant in gauge theories with breaking of conformal invariance. Some well motivated
modifcations of General Relativity where these singularity problems are absent even at the
classical level are discussed. High 
energy density excitations, responsible for
UV divergences in quantum field theories, including quantum
gravity, are likely to be the source of child universes which
carry them out of the original space time. This decoupling could
prevent these high UV excitations from having any influence on
physical amplitudes. Child universe production could therefore be
responsible for UV regularization in quantum field theories which
take into account semiclassically gravitational effects. Child universe
production in the last stages of black hole evaporation, the
prediction of absence of tranplanckian primordial perturbations,
connection to the minimum length hypothesis and in particular the
connection to the maximal curvature hypothesis are discussed. Some discussion of
superexcited states in the case these states are Kaluza Klein excitations is carried out.
Finally, the posibility of obtaining "string like" effects from the wormholes associated
with the child universes is discussed.

\end{abstract}

\maketitle

\newpage

\section{Introduction}

Quantum field theory and quantum gravity in particular suffer from
UV divergences. While some quantum field theories are of the
renormalizable type, quantum gravity is not and the UV divergences
cannot be hidden into a finite number of "counter-terms".
Perturbative renormalizability does not appear to be available for
quantum gravity.

In an apparently unrelated development, the "child universe"
solutions have been studied  \cite{blau}, \cite{ansoldi}. These
child universes are regions of space that evolve in such a way
that they disconnect from the ambient space time. Inflationary
bubbles of false vacuum correspond to this definition \cite{blau},
\cite{ansoldi}. In this case an exponentially expanding
inflationary bubble arises from an ambient space time with zero
pressure which the false vacuum cannot displace. The inflationary
bubbles disconnect from the ambient space generating a child
universe. 

There are difficulties with these
child universe solutions due to their  generic singularity problems \cite{farhiguth} which will very
likely be cured by quantum effects, just in the way that  for example "almost" instanton solutions
are made relevant in gauge theories even with breaking of conformal invariance
(this breaking does not permit the existence of exact classical solutions) \cite{instantons}.
Other avenues for the resolution of the initial singularity problem
of these child universe solution include an initial semiclassical tunneling region that replaces
the singularity \cite{tunneling}, the consideration of violation of energy conditions\cite{conditions}
which itself can originate from quantum effects or the non existence of a Cauchy surface \cite{cauchy}.

There are also some well motivated modifications of General Relativity where singularity problems could be
avoided. For example, in this conference we have heard in the talk by Walter Greiner on his work with Hess on
Pseudo Complex General Relativity \cite{Greiner}, 
which could do this. 

Here we want to explore the possibility that high energy density
excitations, associated to the UV dangerous sector of quantum
field theory could be the source of child universes, which will
carry the UV excitations out of the original ambient space time.
Child universe production could be therefore responsible for UV
softening in quantum field theory that takes into account
gravitational effects. It implies also the existence of a maximum
energy density and curvature.

We will now show now, using a simple model, that very high UV
excitations have appreciable tendency to disconnect from the
ambient space time

\section{The Super High UV Bubble}

We  describe now the model \cite{model} which we will use to describe a high UV
excitation which will be associated with  the production of a
child universe. This model for high UV excitation will consist of
a bubble with very high surface tension and very high value of
bulk energy density inside the bubble.

The entire space-time region consists of two regions and a
boundary: 1) {\bf Region I} de Sitter space 2) {\bf Region II},
Schwarzschild space and  the domain wall boundary separating
regions I and II.

In {\bf Region I}: The de Sitter space. The line element is given
by

\bea ds^2 =-(1-\chi^2 r^2) dt^2 +(1-\chi^2 r^2)^{-1} dr^2 + r^2
d\Omega^2 \label{de} \eea where $\chi$ is the Hubble constant
which is given by \bea \chi^2 = \frac{8}{3} \pi G \rho_0
\label{chi4} \eea

$\rho_0$ being the vacuum energy density of the child universe and
$ G=\frac{1}{m_P^2} $ where $m_P=10^{19}$ GeV.

In {\bf Region II}: The Schwarzschild line element is given by
\bea ds^2 = -(1-\frac{2GM}{r})dt^2 +(1-\frac{2GM}{r})^{-1}dr^2
+r^2 d\Omega^2 \label{ma} \eea

The Einstein`s field equations,  \bea R_{\mu \nu} -\frac{1}{2}
g_{\mu \nu} R =  8 \pi G T_{\mu \nu}. \label{ein} \eea are
satisfied in regions I and II and determine also the domain wall
evolution \cite{blau}, using the methods developed by Israel
\cite{israel}. Using gaussian normal coordinates, which assigns to
any point in space three coordinates
 on the bubble and considers then a geodesic normal to the bubble which
 reaches any given point after a distance $\eta$
 ( the sign of $\eta$ depends on  which side of the bubble the point is
 found). Then energy momentum tensor $T_{\mu \nu}$ is given by \bea
T_{\mu \nu}(x) && = -\rho_0 g_{\mu \nu}, ~~~~~~~~~~~~~~~~(\eta < 0){\rm ~for~ the~ child~ universe, ~~~~{\bf Region ~I}, ~(negative~pressure)} \nonumber \\
&& =0, ~~~~~~~~~~~~~~~~~~~~~~~~(\eta > 0){\rm for~ the~ Schwarzschild ~~region, ~~~~{\bf Region II}} \nonumber \\
&&= - \sigma h_{\mu \nu} \delta(\eta)
~~~~~~~~~~~~~~~~~~~~~~~~~~~~~{\rm for ~ the
~domain~wall~boundary}. \label{tmunu} \eea

where $\sigma$ is the surface tension and $h_{\mu \nu}$ is the
metric tensor of the wall, that is $h_{\mu \nu} = g_{\mu
\nu}-n_{\mu}n_{\nu} $,  $n_{\mu}$ being the normal to the wall.

The eq. of motion of the wall give\cite{blau} \bea M_{\rm cr} =
\frac{1}{2G \chi} \frac{\gamma^3 z_m^6
(1-\frac{1}{4}\gamma^2)^{\frac{1}{2}}}{3 \sqrt{3} (z_m^6
-1)^{\frac{3}{2}}} \label{cr} \eea where the $M_{\rm cr}$ is the
mass at (or above) which there is classically a bubble that
expands to infinity into a disconnected space, the child universe.
In the above equation \bea
&& \gamma = \frac{8 \pi G \sigma }{\sqrt{\chi^2 + 16 \pi^2 G^2 \sigma^2}} \nonumber \\
&& z_m^3 = \frac{1}{2} \sqrt{8+(1-\frac{1}{2}\gamma^2)^2} -
\frac{1}{2} (1-\frac{1}{2}\gamma^2) \label{gm} \eea

where $z^3 = \frac{\chi_{+}^2 r ^3}{2GM}$ and $\chi_{+}^2 = \chi^2
+\kappa^2 $, $\kappa = 4 \pi G \sigma$. $r_m$ is the location of
the maximum of the potential barrier that prevents bubbles with
mass less than $M_{\rm cr}$ to turn into child universes.

We expect this representation of a high UV excitation to be
relevant even for a purely gravitational excitation, which can be
associated , after an appropriate averaging procedure, to an
effective energy momentum, a procedure that gets more and more
accurate in the UV limit.

Let us now focus our attention on the limit where $\sigma
\rightarrow \infty$  (while $\rho_0$ is fixed), which we use as
our first model of a super UV excitation. Then, we see that
$\gamma \rightarrow 2$ and $M_{\rm cr} \rightarrow  0$.
Alternatively, we could obtain another model of super UV
excitation, by considering the energy density inside the bubble,
$\rho_0 \rightarrow \infty$ , while keeping $\sigma$ fixed. This
also leads to $M_{\rm cr} \rightarrow 0$ as well. Finally, letting
both $\sigma \rightarrow \infty$ and $\rho_0 \rightarrow \infty$
while keeping their ratio fixed, also leads to $M_{\rm cr}
\rightarrow  0$. In all these limits we also get the the radius of
the critical bubble $r_m \rightarrow 0$ as well.

In \cite{blau} the above expression for $M_{\rm cr}$ was explored
for the case that energy densities scales (bulk and surface) were
much smaller than the Planck scale, like the GUT scale. This gave
a value for $M_{\rm cr} = 56 kg >>m_p $. Here we take the
alternative view that the scale of the excitations are much higher
than the Planck scale, giving now an arbitrarily small critical
mass. Defining the "scale of the excitation" through by $\rho_0
\equiv M_{exc}^4 $, then the pre-factor $\frac{1}{2G \chi}$ in eq
(\ref{cr}), goes like $\left( \frac{m_p}{M_{exc}} \right)^2 m_p $.
We see that for trans planckian excitations, i.e. if
$M_{exc}>>m_p$, we obtain a very big reduction for $M_{\rm cr}$.
This is a kind of "see saw mechanism", since the higher the
$M_{exc}$, the smaller $M_{\rm cr}$.

This means that in these models for high UV excitations there is
no barrier for the high UV excitation to be carried out to a
disconnected space by the creation of a child universe. Notice
also the interesting "UV -IR mixing" that takes place here:
although we go to very high UV limits in the sense that the energy
density in the bulk or the surface energy density are very high,
the overall critical mass goes to zero.

One should notice that these limits where we take the surface
tension or the energy density to very big values can be achieved
as we go to early times (corresponding to the time of the creation
of the child universes) in models where these quantities are
dynamical variables. In this context \cite{dynten} when
considering for example models with dynamical tension one can show
the existence of child universe production where the critical mass
$M_{\rm cr}$ is indeed zero.

The difficulties related to the singularities for the solutions with
$M > M_{\rm cr}$ should be solved as $M_{\rm cr}$ approaches zero, because then even a very
small quantum fluctuation should be able to wash out this singularity if the mass is very, very small
(the strength of the singularity is associated to the mass of the solution).

\section{The String Gas Shell Example}
Crucial parameters in the child
universe formation models based on the description of vacuum bubbles
in terms of thin relativistic shells are, apart from the total mass--energy
of the asymptotically flat region, the false vacuum energy
density and the shell surface tension. Let us then consider
two spacetime domains ${\mathcal{M}} _{\pm}$ of two
$(3+1)$-dimensional spacetimes, separated by an infinitesimally
thin layer of matter $\Sigma$, a \emph{shell}. We will also assume
spherical symmetry: this simplifies the algebra, and is a
non-restrictive assumption which almost always appears in the
literature. Moreover, as a concrete case \cite{out of nothing} we will choose the one in
which ${\mathcal{M}} _{-}$ is a part of Minkowskii spacetime and
${\mathcal{M}} _{+}$ is a part of Schwarzschild spacetime. Then,
the equations of motion for the shell, i.e. Israel junction
conditions, reduce to the single equation
\begin{equation}
    \epsilon _{-} \sqrt{\dot{R} ^{2} + 1} - \epsilon _{+} \sqrt{\dot{R} ^{2} + 1 - 2 G M / R} = G m (R) / R ,
\label{eq:juncon}
\end{equation}
where $G$ is the gravitational constant and
the only remaining degree of freedom is $R (\tau)$, the
radius of the spherical shell expressed as a function of the
proper time $\tau$ of an observer co-moving with the shell.
$\epsilon _{\pm} = \mathrm{sgn} (n ^{\mu} \partial _{\mu} r
)\rceil _{\mathcal{M} _{\pm}}$ are signs, expressing the fact that
the radial coordinate $r$ can increase ($\epsilon _{\pm} = +1$) or
decrease ($\epsilon _{\pm} = -1$) along the normal direction,
defined by $n ^{\mu} \rceil _{{\mathcal{M}} _{\pm}}$ in
${\mathcal{M}} _{\pm}$, respectively (our convention is
that the normal is pointing from ${\mathcal{M}} _{-}$ to
${\mathcal{M}} _{+}$). The function $m(R)$ is related to the energy--matter
content of the shell, and is what remains of the shell stress--energy
tensor in spherical symmetry after
relating the pressure $p$ and the surface energy density $\rho$ \emph{via} an
equation of state. Let us discuss this point in more detail, since
our choice will be slightly unusual compared to the existing
literature. We will, in fact, use $p = - \rho / 2 $, $p$ being the
uniform pressure and $\rho$ the uniform energy density on
$\Sigma$. A string gas in $n$ spatial dimensions satisfies
$p = - \rho / n $, therefore the two dimensional shell $\Sigma$
we are dealing with is a \emph{sphere of strings}.
This, gives $\rho = \rho_0 / R$, where $\rho_0$ is a
constant, and then, $m (R) = c R$, with $c=4\pi \rho_0$.

Making the above choice, we can then solve the junction condition to obtain the dynamics of the system. It can
be seen that solving (\ref{eq:juncon}) is equivalent to solving the equivalent effective classical problem

\begin{equation}
    \dot{R} ^{2} + V (R) = 0, \quad
    V (R) = 1 - \frac{1}{4 c ^{2}} \left( \frac{2 M}{R} + G c ^{2} \right) ^{2} ,
\label{eq:effequ}
\end{equation}
with the signs determined as $\epsilon _{-} = + 1$ and $\epsilon _{+} = \mathrm{sgn} ( 2 M / R - G c ^{2} )$.
The potential has the following additional properties:
\[\lim _{R \to 0 ^{+}} V (R) =- \infty , \: \lim _{R \to \infty} V (R) = 1 - \frac{G ^{2} c ^{2}}{4}, \:
    \displaystyle \frac{d V (R)}{d R} > 0 .\]
This shows that i) we can have unbounded trajectories
only if $c \geq 2 / G$ and ii) this is independent from the choice of $M > 0$; moreover, from the result for
$\epsilon _{+}$, we have that certainly iii) on all the unbounded trajectories $\epsilon _{+}$ changes
sign (being positive for small enough $R$ and negative for large enough $R$; the general property that
this change of sign must happen behind an horizon, is also obviously satisfied since $c \geq 2 / G$, so that at
$R = 2 G M$ the sign $\epsilon _{+}$ is already negative although for small enough $R$ it is positive).
In view of the global spacetime structure associated with the above properties of \emph{all} the unbounded
solutions , it is clearly seen that they realize the formation of
a baby universe. This happens for a large enough density of strings and \emph{for any} positive value of $M$.
This simple model shows therefore very neatly the phenomena of "child universes out of almost empty space".
At this point we should mention some additional evidence that string matter has some peculiar features related to 
its capability of being responsible to produce universes out of nothing, see for example the interesting arguments of 
Trevisan presented in a  poster in this conference, based on work by Berman and Trevisan \cite{Trevisan}. Another interesting fact
in this respect is that a gas of string matter does not curve the spacetime in the context of the recently formulated
theory with a dynamical time \cite{dyntime}.

\section{The Conjecture}
This allows us to formulate the conjecture that the dangerous UV
excitations that are the source of the infinities and the non
renormalizability of quantum gravity are taken out of the original
space by child universe production, that is, the consideration of
child universe production in the ultrahigh (trans planckian)
sector of the theory could result in a finite quantum gravity,
since the super high UV modes, after separating from the original
space will not be able to contribute anymore to physical
processes.

The hope is that in this way, child universes could  be a of
interest not only in cosmology but could become also an essential
element necessary for the consistency of quantum gravity.  One
situation where all the elements required (high energy densities ,
since the temperature is very big) necessary for obtaining a child
universe appears to be the late stages of Black Hole evaporation.
If the ideas explained here are correct, we should not get
contributions to primordial density perturbations from the trans
planckian sector, since these perturbations would have
disconnected from our space time. Also, any attempt to measure
distances smaller than the Planck length will be according to this
also impossible since such a measurement will involve exciting a
high UV excitation that will disconnect. This means that there
must be a minimum length that we could measure, of the order of
the Planck scale.

It appears there is a maximal energy density according to this,
since now bubbles with high energy density will be quickly
disconnected, being replaced  in the observable universe by
regions of Schwarzschild space, which has zero energy density,
i.e., a very big energy density must decay in the observable
universe. The "maximal curvature"\cite{maxcurv} hypothesis (here
we focus on scalar curvature) is justified by this maximal energy
density result, if we use eq. (\ref{ein}). An effective dynamics
that takes into account the effect of child universe production
(i.e. integrates out this effect) could resemble indeed that of
obtained using the maximal curvature hypothesis
\cite{maxcurv}. 

\section{Super Excited Kaluza Klein (KK) Modes}
The gravitational effects on the spectrum of modes of particles with momentum in the direction 
of some periodic dimension have been analyzed \cite{graveffects}. It was found there that the naive,
uniformly spaced KK excitation spectrum gets drastically modified, since the size of the compact dimentions
likes to grow near the region where the KK excitation lives. This leads to many
orders of magnitude decrease of the energy of these KK excitations. The modification of the
spectrum of KK excitations due to the growing of the extra dimensions was studied also in \cite{Morris}.
Furthermore, once a configuration like that has been created, it is energetically possible for
these modes to become superheavy by the recolapse of the extra dimension provided a child universe 
with an associated wormhole region is created \cite{univcreation} . The superheavy modes must then decouple
from the ambient universe for this to happen, in agreement with our general picture.

\section{Child Universes, Wormholes and Strings}
The creation of a child universe implies the creation of a wormhole region.
Static wormhole configurations have been studied since the first construction
by Einstein and Rosen \cite{ER}, which consisted of simply joining two exterior Schwarzschild at
the horizon, producing a doubling of the space for $r>2M$, but the elimination of the space $r<2M$. 
Although not noticed by Einstein and Rosen, the consistency
of this construction (that is in order for it to be a solution of Einstein`s equations) requires
the existence of light like matter at $r = 2M $ \cite{GKNP}.
Generally wormholes are considered by joining exterior solutions outside the horizon $r>2M$, for a review
see  \cite{Visser}. Tranversable wormholes require in general exotic matter.
Electric field lines can go from one universe to the other going through the wormhole and causing the appearence
of charge, positive appearence on one universe and negative appearence on the other universe as has been pointed out
by Misner and Wheeler \cite{MisnerWheeler}. 

The gauge fields that go from one side of the wormhole to the other can be used to construct wormhole 
throats which can be very, very long \cite{Guendelworm}, \cite{Dz}. Then it so happens that
 these very long wormhole 
throats have a dynamics that mimics that of string theory\cite{Dz}, which raises the interesting question of whether wormhole theory
and in particular child universe theory could be origin of string theory, which could appear as the effective description of
the dynamics of these long wormhole throats.

\section*{Acknowledgments} 
I want to thank the organizers of IWARA2009 conference for inviting me to this very interesting event and for support, to  Vladimir Dzhunushaliev for reading the manuscript and for interesting comments, to Walter Greiner for interesting conversations concerning  avoidance of singularities in his model and relevance to a child universe production, to Marcelo Samuel Berman and Luis Augusto Trevisan for interesting conversations concerning the possibility of creating a universe out of nothing.  This manuscript was prepared while I was visiting the Astrophysics and Cosmology Group at the Pontificia Universidad Catolica de Valparaiso, Chile.

\end{document}